\documentclass[preprint,times,sort&compress]{elsarticle}

\usepackage{graphicx}
\usepackage{natbib,geometry,fleqn,txfonts,hyperref}
\usepackage{xcolor}

\journal{Acta Meterialia}
	\newcommand{\figref}[1]{Fig.~\ref{#1}}

	\newcommand{\ATO}{ATiO${}_3$}
	\newcommand{\BTO}{BaTiO${}_3$}
	\newcommand{\PTO}{PbTiO${}_3$}
	\newcommand{\STO}{SrTiO${}_3$}
	
	\newcommand{\BCTO}{Ba${}_{1-x}$Ca${}_{x}$TiO${}_3$}
	\newcommand{\BZT}{BaTi${}_{1-x}$Zr${}_x$O${}_3$}
	\newcommand{\SRO}{SrRuO${}_3$}
	\newcommand{\LSAT}{(LaAlO${}_3$)${}_{0.3}$-(SrAl${}_{0.5}$Ta${}_{0.5}$O${}_3$)${}_{0.7}$}
	\newcommand{\TiOct}{TiO${}_6$}

	\newcommand{\ttg}{$t_{\mathrm{2g}}$}
	\newcommand{\eg}{$e_{\mathrm{g}}$}

	\newcommand{\ang}[1]{#1${}^\circ$}
	\newcommand{\um}{$\mathrm{\mu m}$}
	\newcommand{\us}{$\mathrm{\mu s}$}
	\newcommand{\uA}{$\mathrm{\mu A}$}
	
	\newcommand{\etal}{\textit{et~al.}}
	\newcommand{\dti}{$\delta_\mathrm{Ti}$}
	\newcommand{\tw}{$\tau_w$}

\begin{document}

\begin{frontmatter}
	\title{Dielectric response of \BTO{} electronic states under AC fields via microsecond time-resolved X-ray absorption spectroscopy}

	\author[HU]{S.~Kato\corref{mycorrespondingauthor}}
	\cortext[mycorrespondingauthor]{Corresponding author}
	\ead{kato-seiya@hiroshima-u.ac.jp}

	\author[HU]{N.~Nakajima}
	\author[TIT_material,TIT_nuclear]{S.~Yasui}
	\author[TIT_material,TIT_chem]{S.~Yasuhara}
	\author[SU_elec,SU_eng,SU_opt]{D.~Fu}
	\author[KEK]{J.~Adachi}
	\author[KEK]{H.~Nitani}
	\author[KEK]{Y.~Takeichi}
	\author[LU]{A.~Anspoks}

	\address[HU]{Graduate School of Advanced Science and Engineering, Hiroshima University, 1-3-1 Kagamiyama, Higashihiroshima, Hiroshima 739-8562, Japan}
	\address[TIT_material]{Laboratory for Materials and Structures, Tokyo Institute of Technology, 4259-J2-19 Nagatsuta-cho, Midori-ku, Yokohama 226-8503, Japan}
	\address[TIT_nuclear]{Laboratory for Advanced Nuclear Energy, Tokyo Institute of Technology, 2-12-1 N1-16 Ookayama, Meguro-ku, Tokyo 152-8550, Japan}	
	\address[TIT_chem]{School of Materials and Chemical Technology, Tokyo Institute of Technology, 2-12-1 Ookayama, Meguro-ku, Tokyo 152-8550, Japan}
	\address[SU_elec]{Department of Electronics and Materials Science, Faculty of Engineering, Shizuoka University, 3-5-1 Johoku, Naka-ku, Hamamatsu 432-8561, Japan}
	\address[SU_eng]{Department of Engineering, Graduate School of Integrated Science and Technology, Shizuoka University, 3-5-1 Johoku, Naka-ku, Hamamatsu 432-8561, Japan}
	\address[SU_opt]{Department of Optoelectronics and Nanostructure Science, Graduate School of Science and Technology, Shizuoka University, 3-5-1 Johoku, Naka-ku, Hamamatsu 432-8561, Japan}
	\address[KEK]{Institute of Materials Structure Science, High Energy Accelerator Research Organization (KEK), 1-1 Oho, Tsukuba, Ibaraki 305-0801, Japan}
	\address[LU]{Institute of Solid State Physics, University of Latvia, Kengaraga street 8, Riga, LV-1063, Latvia}

	\begin{abstract}
		For the first time, the dielectric response of a \BTO{} thin film under an AC electric field is investigated 
		using microsecond time-resolved X-ray absorption spectroscopy at the Ti K-edge 
		in order to clarify correlated contributions of each constituent atom on the electronic states.
		Intensities of the pre-edge \eg{} peak and shoulder structure just below the main edge 
		increase with an increase in the amplitude of the applied electric field,
		whereas that of the main peak decreases in an opposite manner.
		Based on the multiple scattering theory, 
		the increase and decrease of the \eg{} and main peaks are simulated for different Ti off-center displacements. 
		Our results indicate that these spectral features reflect the inter- and intra-atomic hybridization of Ti 3$d$ with O 2$p$ and Ti 4$p$, respectively.
		In contrast, the shoulder structure is not affected by changes in the Ti off-center displacement 
		but is susceptible to the effect of the corner site Ba ions.
		This is the first experimental verification of electronic contribution of Ba to polarization reversal.
	\end{abstract}


\end{frontmatter}
	\section{Introduction}
		Ferroelectric materials are widely used in various practical applications,
		including in multilayer ceramic capacitors, actuators, and memory cells, among others.
		However, to understand the dielectric response of these materials,
		it is essential to first understand the dynamic behavior of spontaneous polarization reversal under an electric field.
		Among ferroelectric materials, perovskite titanates (\ATO{}; A=Pb, Ba, Sr, Ca) attract significant attention
		owing to the versatility of the functional solid solutions that are composed from them. 
		In particular, a lack of space-inversion symmetry is a prerequisite for the occurrence of spontaneous polarization; 
		therefore, previous related literature is primarily focused on the crystal structure.
		In the case of perovskite titanates, the off-center displacement of Ti ions in the \TiOct{} octahedra
		is a major cause of polarization.~\cite{Devonshire, Deguchi, Nakajima_2010, Kawakami_2013, Kawakami_2015, Lu_2017} 
		Various studies have been performed to study X-ray absorption of Ti K-edge for \BTO{}.~\cite{Ravel_1995, Ravel_1998, Bootchanont_2014, Phaktapha_2017, Yoshiasa_2018}

		Furthermore, it is obvious that also A-site cations significantly affect dielectric properties of \ATO{},
		including magnitude of polarization and Curie temperature.~\cite{databaseBTO, databaseSTO, databasePTO}
		Cohen was the first to theoretically demonstrate the existence of Pb 6$s$ and O 2$p$ hybridization in \PTO{},~\cite{Cohen}
		going beyond the hard-sphere model.
		This prediction was experimentally verified by Kuroiwa \etal{} via high-precision X-ray diffraction analysis, 
		which revealed the electron density distribution between Pb and apical O.~\cite{Kuroiwa}
		Recently, Anspoks \etal{} reported the correlation effects of A-site ions in O and Ti based on the 
		reverse Monte Carlo method applied to the extended X-ray absorption fine structure (EXAFS) data.~\cite{Anspoks} 
		
		Thus, it is quite evident that orbital hybridization between constituent atoms is closely linked to the dielectric properties of \ATO{}.
		Hence, direct observation of the electronic states under the application of an electric field is an appropriate approach
		to investigate the dielectric properties of ferroelectric materials.
		We employed microsecond time-resolved (TR) X-ray absorption spectroscopy (XAS)
		because tiny changes in spectra are expected by the previous studies.~\cite{Moriyoshi, Tazaki}
		The TR measurement is essential to exclude any undesired influences caused by a DC measurement such as Joule heating and fatigues.
		In this study, we selected \BTO{} as the target material
		whose electronic states at the instant of polarization reversal were observed
		via microsecond TR-XAS measurements. 
		XAS is an element-specific technique, which is sensitive to local bonding around absorbing atoms.
		Combined with the TR approach,
		XAS can be used for the investigation of the polarization reversal response of the atomic bonds in \BTO{}.

		In particular, in this study, TR-XAS data of \BTO{} at the Ti K-edge are used
		to investigate the response of the corresponding electronic states under an applied AC electric field. 
		It was observed that intensities of the pre-edge \eg{} peak and shoulder structure just below the main edge
		increase as the electric polarization increases, whereas that of the main peak decreases in an opposite manner.
		This observation indicates that in addition to Ti 3$d$-O 2$p$ hybridization,
		a correlated effect of Ba ions on Ti electronic states also contributes to polarization reversal in \BTO{}. 

	\section{Experimental}
\begin{figure}
			\includegraphics{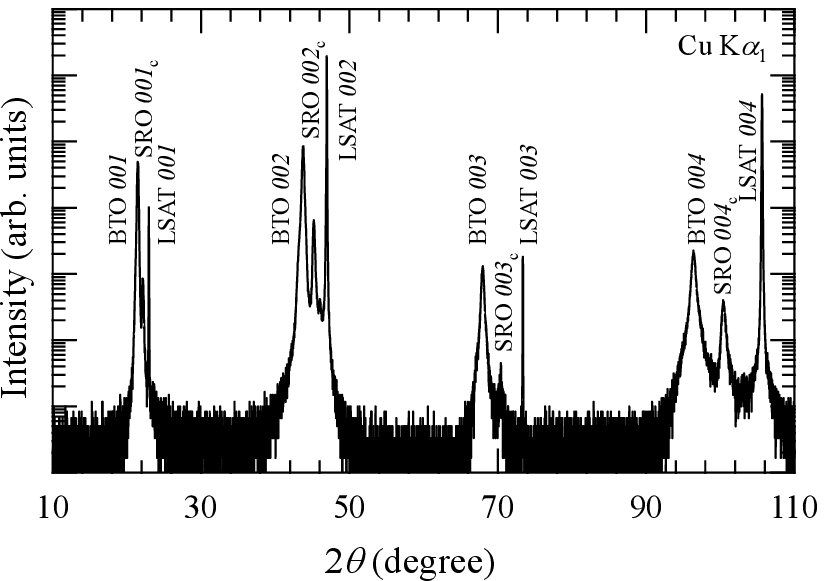}%
			\caption{X-ray diffraction pattern of the \BTO(001) thin film grown on a \SRO{}(001)/\LSAT{}(001) substrate.}
			\label{fig:XRD}
		\end{figure}

		An epitaxial \BTO{}(001) film with a thickness of 650~nm was prepared using pulsed laser deposition
		on a 0.5~mm thick \LSAT{}(001) substrate along with a 100~nm thick \SRO{}(001) buffer layer,
		which also acted as the bottom electrode.
		A 50~nm thick Pt with a diameter of 100~\um{} was evaporated on the \BTO{} film, which acted as the top electrode.
		Film growth was verified via X-ray diffraction as shown in \figref{fig:XRD}. 
		It was confirmed that all layers had a preferred (001) orientation normal to the surface of the film.
		In addition, the lattice constant, $c$, of the \BTO{} film calculated based on the (002) peak was 4.136~\AA{}, 
		which is longer than that of the standard powder ($c$ = 4.018~\AA).~\cite{databaseBTO}
		This lattice mismatch between the \BTO{} film and substrate induces a compressive strain.~\cite{Choi}
		The preferred orientation of $c$-axis is also reported in thin films with \STO{} substrates where misfit strain is smaller than our sample.~\cite{Tyunina} 
		This result guarantees that the influence of $a$ domain and the domain wall motion under electric fields are negligible in our case.

		\begin{figure}
			\includegraphics{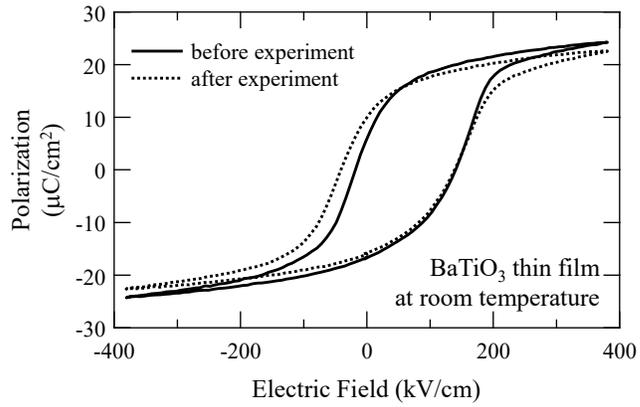}%
			\caption
			{
				$P$-$E$ hysteresis loops of the \BTO{} thin film before (solid) and after (dotted) TR-XAS measurements.
			}
			\label{fig:hys}
		\end{figure}%

		The ferroelectricity of the film was verified by measuring polarization-electric field ($P$-$E$) hysteresis
		using a ferroelectric tester (Toyo Corporation, FCE-fast) with a triangular electric field of 1~kHz;
		these results are shown in \figref{fig:hys}.
		The open loops indicate ferroelectricity of the film both before (solid) and after (dotted) TR-XAS measurements.
		These loops are shifted in the positive electric field direction indicating a preferred downward polarization.~\cite{Setter, Shin} 
		This imprint effect is due to the extrinsic interface effect,
		including the asymmetric properties of the top and bottom electrodes.~\cite{Arlt}
		It should be noted that whereas the saturation polarization decreased,
		the magnitude of the negative coercive field increased after the TR-XAS measurements;
		these observations might be attributed to the fatigue effects caused by domain pinning and microcracking;~\cite{Luo}
		however, the dielectric properties of the sample remain even after experiments because hysteresis loop is closed as shown in \figref{fig:hys}, 
		indicating low leakage current density. 
		The currents at the maximum voltage before and after the TR-XAS measurements were 1.1~\uA{} and 1.2~\uA{}, respectively.
		The difference is below the experimental error, therefore it does not produce any detectable change in the TR-XAS data.
		
\begin{figure}
			\includegraphics{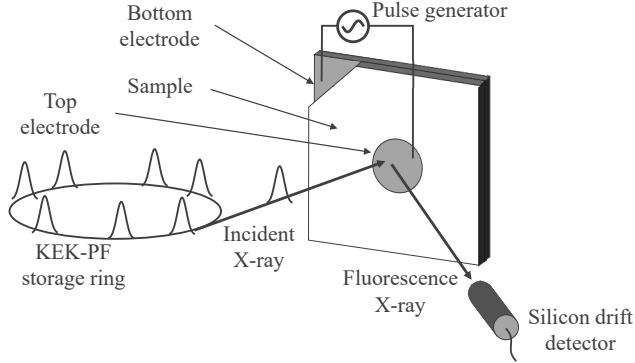}%
			\caption{Schematic diagram of the TR-XAS experimental setup.}
			\label{fig:setup}
		\end{figure}

		XAS experiments were performed on the beamline BL-15A1 equipment at the Photon Factory of the High Energy Research Organization (KEK-PF)
		using a Si (111) liquid-nitrogen cooled double-crystal monochromator.
		The X-ray beam was focused to 20~\um{}(H) $\times$ 20~\um{}(V) at the sample,
		which ensured the inclusion of the entire beam on a top electrode.
		Ti K-edge spectra were recorded in partial fluorescence yield mode using a Si drift detector (SDD), 
		as illustrated in the schematic diagram shown in \figref{fig:setup}.
		X-rays were incident on the top electrode of the film at an angle of \ang{45} from the surface normal.
		Furthermore, the detected X-rays were converted to digital signals
		using a digital signal processor (DSP; Techno-AP, APU101) with the addition of a time stamp for each signal.
		
		\begin{figure}
			\includegraphics{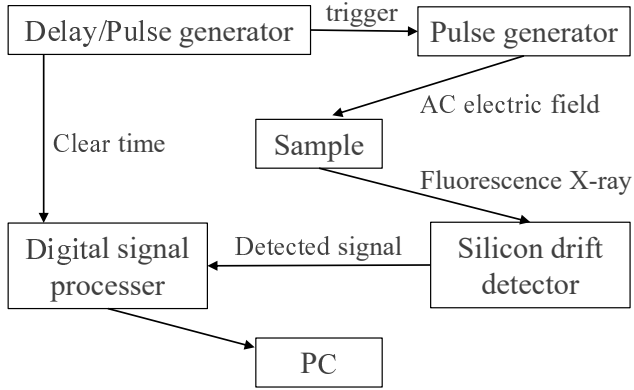}%
			\caption{Block diagram of the signal processing system synchronized with the applied voltage.}
			\label{fig:tr_block}
		\end{figure}%
		
		TR-XAS measurements were realized using the time-stamp information based on the step-scan method. 
		\figref{fig:tr_block} shows the block diagram of the signal processing system employed in this study. 
		The internal time of the DSP was repeatedly reset to 0 every 3~ms
		using a clear pulse generated by a delayed pulse generator (SRS, DG645).
		In addition, the same clear pulse was also input into the wave generator as a trigger pulse	to apply the electric field to the film.
		In particular, one cycle of the applied voltage consisted of a single triangular pulse ($\pm$25~V, 1~kHz) and null voltage for 2~ms.
		Using this cycle, it was possible to assign a timing to the detected X-ray signals in response to the applied voltage,
		thereby enabling TR measurements.
		Several scans were summed to obtain the final Ti K-edge spectra with sufficient statistics.
		Because TR-XAS measurements require a long duration---e.g., 8~h for one spectral region--- 
		we focused on only three spectral features in our study, namely \eg{} peak, shoulder structure, and main peak.

	\section{Results \& Discussion}
\begin{figure}
			\includegraphics{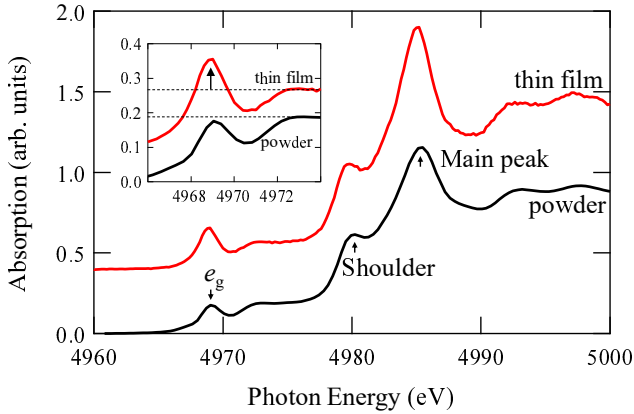}%
			\caption{Ti K-edge X-ray absorption near-edge spectra for a \BTO{} thin film and standard powder at room temperature.}
			\label{fig:staticXAS}
		\end{figure}

		A non-TR Ti K-edge spectrum of the \BTO{} thin film is shown in \figref{fig:staticXAS}, 
		along with that of standard \BTO{} powder as a reference. 
		The absorption profile above the sharp main peak at 4985~eV corresponds to the density of states of the unoccupied Ti 4$p$ states, 
		whereas small features in the pre-edge region (4965--4975~eV) represent the Ti 3$d$ states.
		In the case of an octahedral symmetry, 
		the five-fold degenerate 3$d$ orbitals split into three-fold and two-fold degenerate states with \ttg{} and \eg{} symmetries, respectively. 
		A tiny hump at 4967~eV corresponds to the \ttg{} peak, whereas clear peaks at 4969~eV can be attributed to \eg{}.	
		The reason why \ttg{} peak intensity is very weak is mainly because it is forbidden in a dipole and is allowed in quadrupole approximation. 
		In the K-edge spectra, the probability of the 1$s$ to 3$d$ transition is much lower than that of the dipole transition. 
		This effect can be observed in other bulk materials.~\cite{Vedrinskii, Groot} 
		The intensity of the \eg{} peak increases as the local distortion in the \TiOct{} octahedron increases, 
		because the $pd\sigma$ hybridization between the Ti-3$d$ \eg{} orbitals and O $p\sigma$ orbitals becomes more pronounced, 
		resulting in a relatively large dipole component.
		It has been reported that the intensity of the \eg{} peak is proportional to the mean-square displacement 
		of a Ti ion from the center of the \TiOct{} octahedron.~\cite{Vedrinskii, Anspoks_2014} 

		As shown in the inset of \figref{fig:staticXAS}, the \eg{} peak of the \BTO{} thin film is larger than that of standard powder.
		In addition, the \ttg{} hump of the thin film almost flattens out.
		This observation can be attributed to the compressive strain
		resulting from the lattice mismatch between the \BTO{} film and underlying substrate.
		Because the \BTO{} film has a preferred orientation of the $c$-axis with enhanced tetragonality,
		a distinct dielectric response can be expected.
		
\begin{figure}
			\includegraphics{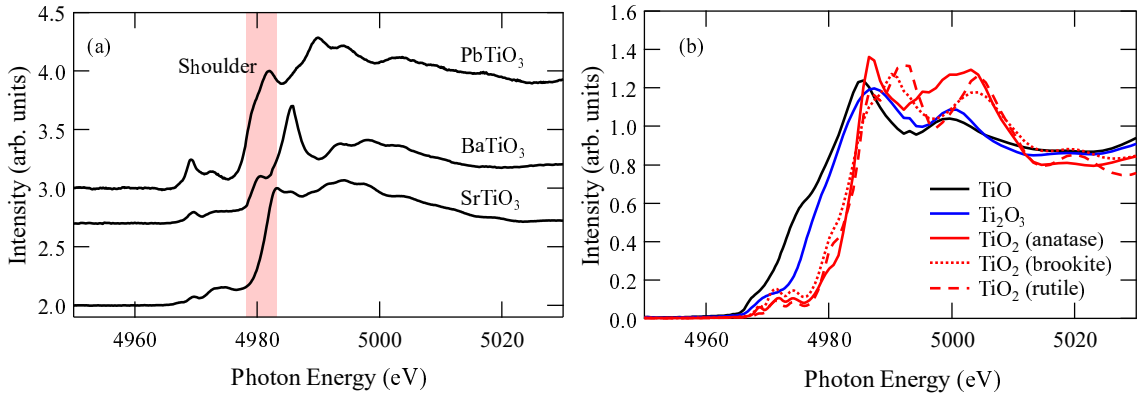}%
			\caption
			{
				Ti K-edge X-ray absorption near-edge spectra of (a) \ATO{} powders (A = Pb, Ba, and Sr)
				and (b) TiO, Ti$_2$O$_3$, and TiO$_2$ powders.
			}
			\label{fig:xasATO}
		\end{figure}

		The shoulder structure at 4980~eV is a characteristic of \BTO{}.
		Though the electronic states responsible for this structure have not been identified yet,
		the shoulder structure indicates the contribution of the A-site ions in the ferroelectric perovskite \ATO{};
		the room-temperature spectra of commercial \ATO{} powders (A = Pb, Ba, and Sr) are shown in \figref{fig:xasATO}(a).
		XAS technique is sensitive to the symmetry and electronic states.
		As the electronic states differ according to the chemical composition, so do the absorption spectra.
		For example, the Ti-K edge spectra of TiO, Ti$_2$O$_3$, and TiO$_2$ (anatase, rutile, and brookite) shown in \figref{fig:xasATO}(b) are different  
		because their electronic states such as band structures are completely different.

		In contrast to paraelectric \STO{} with the cubic symmetry for which no shoulder structure is observed in its spectra, 
		ferroelectric \BTO{} and \PTO{} with the tetragonal symmetry do exhibit shoulder structures. 
		Differences in near edge structure between \BTO{} and \PTO{} may be due to differences in the hybridization 
		of Ti 4$p$ states with the states of A-site ions (i.e., Ba 6$p$ vs Pb 6$p$).~\cite{Blanchard}
		Moreover, the intensity of these structures seems to depend on the degree of ferroelectricity;
		accordingly, \PTO{} has a more prominent shoulder structure compared with \BTO{}.

\begin{figure}
			\includegraphics{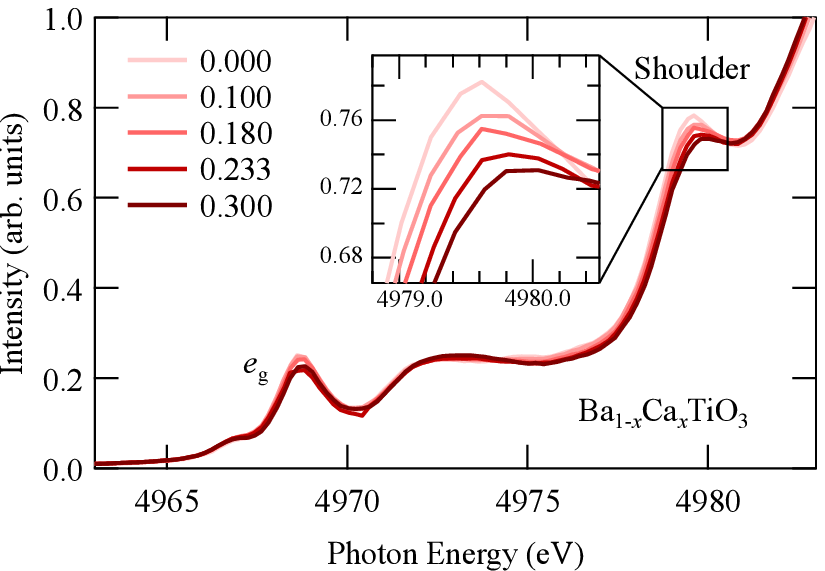}%
			\caption
			{
				Ti pre-edge X-ray absorption spectra of \BCTO{} powders ($x=$0, 0.1, 0.18, 0.233, and 0.3).
			}
			\label{fig:xasBCTO}
		\end{figure}

		This trend can also be confirmed based on a series of Ti K-edge spectra of \BCTO{} ($x =$ 0, 0.1, 0.18, 0.233, and 0.3),
		which are shown in \figref{fig:xasBCTO}.
		In particular, \BCTO{} stays in the tetragonal phase within the considered $x$ range,
		and therefore, ferroelectricity remains stable.~\cite{Fu, Fu2}
		As can be seen in \figref{fig:xasBCTO}, the intensity of the shoulder structure decreases monotonically with increasing $x$,
		i.e., decreasing Ba concentration.
		Similar results have been reported for \BZT{}.~\cite{Levin, Bootchanont}
		Likewise, the intensity of the shoulder structure decreases with an increase in $x$,
		i.e., decrease in the Ti content.
		From these results, it can be concluded that the intensity of the shoulder structure in the \BTO{} spectra reflects
		the electronic hybridization between Ti and Ba.
		This seemingly implausible claim has a theoretical background
		proposed by Ghosez \etal{} based on the Born effective charges of each constituent atom.~\cite{Ghosez}

\begin{figure}
			\includegraphics{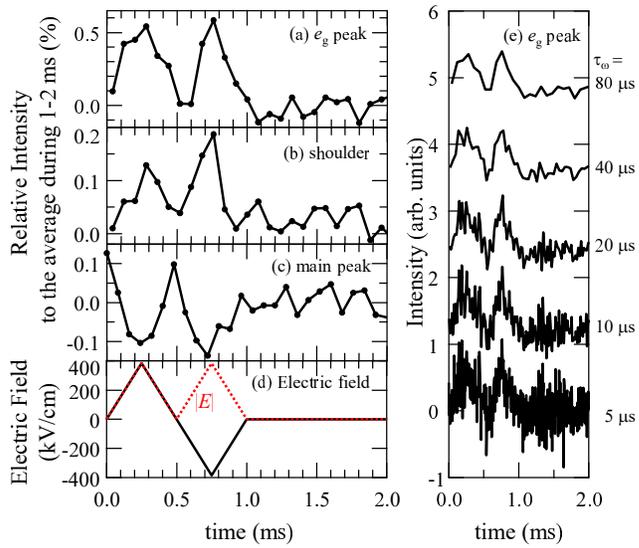}%
			\caption
			{
				Time variation of the intensities of the (a) \eg{} peak, (b) shoulder structure, 
				and (c) main peak under the application of (d) a triangular AC electric field.
				(e) Temporal variation of the \eg{} peak for various time windows \tw{}.
			}
			\label{fig:I-t}
		\end{figure}

		As discussed above, because the \eg{} peak and shoulder structure directly reflect
		the ferroelectricity of perovskite titanates,
		we focused on these structures as well as the main peak in our TR experiments. 
		The time variation of the applied electric field ($E$) for one cycle is shown in \figref{fig:I-t}(d);
		a symmetric positive-and-negative triangular field for 1~ms
		and a field-free duration time for 1~ms were applied in sequence.
		The time variation of the triangular pulse corresponds to 1~kHz, 
		which is the same frequency as that used for the hysteresis measurements.
		It should be noted that the field-free region is included to prevent Joule heating as well as to provide the field-free condition.
		In addition, we utilized not static but AC electric field to eliminate any undesired influence from fatigue, 
		which ensures that the relative change in a short time frame is detectable.
		We have not observed any measurable drift of the observed values upon time.

\begin{figure}
			\includegraphics{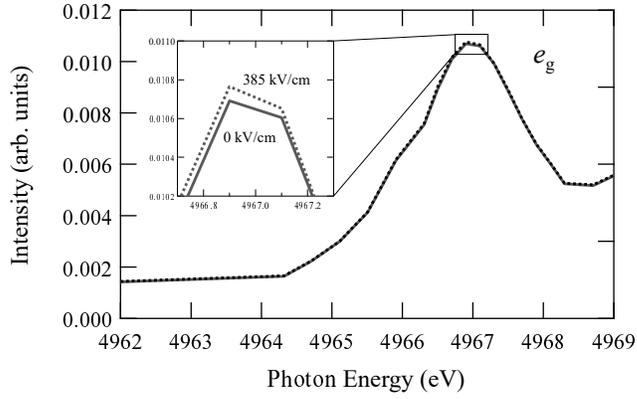}%
			\caption
			{		
				Ti pre-K edge X-ray absorption spectra of \BTO{} with (solid) and without (dashed) electric fields.
			}
			\label{fig:eg_withV}
		\end{figure}

		The maximum variation of spectra with electric fields is shown in \figref{fig:eg_withV}.
		The increase of \eg{} peak intensity is 0.6~\% and the peak shifts are not observed.
		These tiny changes are due to a small distortion caused by electric fields.~\cite{Tazaki, Moriyoshi} 
		The unchanged Ti oxidation state before and after experiments is verified by seeing peak positions in XAS spectra. 
		If the valence number of Ti or its local structure changes, the chemical shifts will occur as it is seen in spectra of TiO, Ti$_2$O$_3$, and TiO$_2$,
		which are presented in \figref{fig:xasATO}(b).
		In our case, we did not observe any changes in the peak positions with and without electric fields.
		Then, we focus on the time variation of intensities.

		The time variation of the area-integrated intensities of the \eg{} peak, shoulder structure, 
		and main peak are presented in \figref{fig:I-t}(a), (b), and (c), respectively. 
		The integration area is set to $\pm$1~eV around the local maxima of each spectral feature,
		and the integrated intensity is normalized to unity by the average intensity around an applied field of 0 V for 1--2~ms.
		The width of the time window (\tw{}) for each plot is 80~\us{}.
		The signal to noise ratio of intensities can be estimated in the region from 1--2~ms.
		Intensities in this region should be constant but there is still noise left. 

		It is obvious that the intensities of the \eg{} peak and shoulder structure are synchronized with 
		the magnitude of the applied electric field $|E|$ beyond the noise level of $\sim$0.05\%.
		Intensities of both features increase as the magnitude of the applied field increases. 
		The interaction of X-rays with the electronic states of \BTO{} is dependent not on the polarity of 
		the electric polarization but on the magnitude of polarization;
		therefore, one-half period of the applied field was observed. 
		This result confirms that not only the \eg{} peak
		but also the shoulder structure is associated with the ferroelectricity of \BTO{}.
		In contrast, the intensity of the main peak changes in the opposite manner, i.e., its intensity decreases with increasing $|E|$. 
		A naive interpretation of this opposite trend is the broadening of the 4$p$ electronic states
		because of the reduction in the coordination symmetry of the Ti ion under the applied electric field.
				
		To better understand the time variation of the spectral features,
		we studied the data presented in \figref{fig:I-t}(e).
		In particular, this figure illustrates the time variation of the \eg{}-peak intensity for various \tw{}.
		Because we detect all SDD signals using the time stamp information,
		it is possible to change the \tw{} to any period longer than the minimum interval of the DSP-processed signals, 
		which is typically around 0.1~\us{} and depends on the setting of the instrument.
		The time variation of the \eg{}-peak intensity is evident at any \tw{},
		even though the noise level increases as the \tw{} becomes smaller.
		In general, the response of the electronic states to the applied electric field is instantaneous; 
		therefore, time variation of the spectral features shows the same trend regardless of \tw{}.
		
\begin{figure}
			\includegraphics{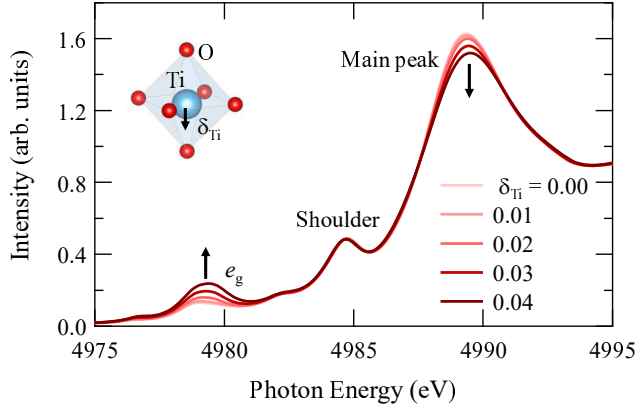}%
			\caption
			{
				Simulated Ti K-edge spectra of \BTO{} for various Ti off-center displacements 
				(\dti{}) from \dti{} to 0.04 in atomic units. 
				Calculations were performed using the FEFF 9.6 program.
			}
			\label{fig:FEFF_Ti_off}
		\end{figure}

		To provide a theoretical background for the interpretation of the observed spectral features, 
		the experimental spectra were compared with the simulated spectra obtained using the FEFF 9.6 program, 
		which is based on the multiple scattering theory.~\cite{FEFF} 
		The simulation results are shown in \figref{fig:FEFF_Ti_off}, wherein the Ti off-center displacement (\dti{}) 
		was varied from \dti{}=0 (body-center) to 0.04 in atomic units. 
		It was assumed that larger displacements would make the spectral changes clear. 
		As already discussed in the literature,~\cite{Vedrinskii} the intensity of the \eg{} peak increases with increasing \dti{}, 
		which reflects the enhanced Ti 3$d$ \eg{}-O 2$p$ hybridization. 
		In contrast, the intensity of the main peak decreases in an opposite manner.
		The simulated results suggest that not only the effect of broadening the Ti 4$p$ states
		but also the compensation for the number of unoccupied states would be a plausible reason
		for the observed spectral features.

		In contrast to the abovementioned two features,
		the shoulder structure is less sensitive to the Ti off-center displacements, 
		which clearly indicates a different physical origin for the shoulder structure. 
		As previously mentioned,
		the shoulder structure was attributed to the A-site contribution in Ti as per several experimental results.
		One possible explanation for the A-site contribution is the Born effective charge.
		The Born effective charge tensor, $Z_{\kappa, \gamma \alpha}^{*}$, of an atom $\kappa$
		can be linked to the change of polarization $P_\gamma$
		induced by the periodic displacement $\tau_{\kappa, \alpha}$
		along the $\alpha$ direction induced on the atom $\kappa$ by an external field
		$\varepsilon_\gamma$: $Z_{\kappa, \gamma \alpha}^{*}=V\partial{P_\gamma}/\partial{\tau_{\kappa, \alpha}}$,
		where $V$ is a unit cell volume.~\cite{Ghosez, Tulip}
		For the simplest case, if we consider a single dipole model, the Born effective charge $Z^*$ can be expressed by
		a static charge $Z$ by using an interatomic distance $u$ as $Z^{*}(u)=Z(u)+u\partial{Z(u)}/\partial{(u)}$.
		If $Z(u)$ changes rapidly with $u$, the difference between $Z(u)$ and $Z^{*}(u)$ becomes large.
		Ghosez \etal{} concluded that $Z_{\mathrm{Ba}, \mathrm{33}}^{*}$ and $Z_{\mathrm{Ti}, \mathrm{33}}^{*}$,
		the Born effective charge of Ba and Ti, respectively, along the ferroelectric polarization $P_3$,
		are comparable to the charge transfer between Ti and apical O along the polarization direction.
		Therefore, it can be stated that orbital hybridization is not restricted to Ti and O but also involves Ba,
		or at least, that Ba also plays an important role in the formation of the valence band.

		The Born effective charge makes a major contribution to the formation of the shoulder structure;
		the term $u\partial{Z(u)}/\partial{u}$ is essential for the covalent character of the Ti-O bond.
		This term would be insensitive to the static Ti off-center displacement, as shown in the simulation in \figref{fig:FEFF_Ti_off}.
		Under the application of an AC electric field, not only Ti ions but also Ba ions are perturbed
		leading to the non-negligible enhancement of $u\partial{Z(u)}/\partial{u}$ as compared to the static environment.
		The intensity of the shoulder structure increases in phase with that of the \eg{} peak
		under the application of the AC field (\figref{fig:I-t}).
		Thus, it can be speculated that the charge transfer from the Ti 4$p$ state to the Ba unoccupied state,
		e.g. the 5$d$ state, occurs as well as the charge transfer from the Ti 3$d$ state to the O 2$p$ state,
		as concluded from the \eg{} peak.~\cite{Vedrinskii}
		The shoulder structure of the Ti K-edge spectrum of \BTO{} provides
		an electronic point of view for the dynamic piezoelectric effect of \BTO{}.
		
		Furthermore, though the A-site contribution to the ferroelectric properties of \ATO{}
		has been observed and studied in many previous works,
		the polarization response under applied AC electric fields using XAS measurements
		provides a new insight into this, especially in terms of energetics of electronic hybridization.
		In this light, Ba K-edge EXAFS would provide more advanced information on the A-site contribution
		to the ferroelectric property of \ATO{};
		therefore, another TR-XAS experimental study using \BTO{} under applied electric fields is currently in progress.
		
	\section{Conclusions}
		For the first time, Ti K-edge XAS measurements for a \BTO{} thin film were performed
		under the application of a triangular electric field in the TR mode.
		This technique enables us to detect tiny changes in spectra and observe electronic states under electric fields.
		Three characteristic features were studied, namely the pre-edge \eg{} peak, shoulder structure, and main peak.
		Among these three features, the responses of the first two to the electric field were in phase,
		whereas that of the latter was antiphase.
		Using model simulations, it was confirmed
		that Ti off-center displacements affect the increase in the \eg{}-peak intensity and decrease in the main peak intensity. 
		However, it was observed that the shoulder structure was not influenced by the Ti off-center displacement.
		Nevertheless, it showed a dielectric response to the AC electric field
		that can be consistently explained from the perspective of the Born effective charge.
		Thus, the Ba contribution to the ferroelectric property of \BTO{} via electronic hybridization
		is concluded experimentally using the novel TR-XAS technique.
		The TR-XAS technique's utilization of the full time stamp information of the DSP is versatile for many applications;
		therefore, it provides a new approach for the study of the dynamical behavior of electronic states.
		
	\section*{Acknowledgements}
		This research was performed under the approval of the Photon Factory Program Advisory Committee 
		(PF-PAC; Contract Numbers 2015G580, 2017G587, and 2019G614) and was financially supported by 
		JSPS KAKENHI Grant Numbers 18H01153, 19H02426, and 18K19126.
		The experiment for measuring spectra in \figref{fig:xasATO}(b)
		was performed on beamline BM26A (proposal MA  2731) at the European Synchrotron Radiation Facility (ESRF), Grenoble, France. 
		We are grateful to Local Contact at the ESRF for providing assistance in using beamline BM26A.
		Institute of Solid State Physics, University of Latvia as the Center of Excellence has received funding from the European Union's Horizon 2020 Framework Programme H2020-WIDESPREAD-01-2016-2017-TeamingPhase2 under grant agreement No.~739508, project CAMART${}^2$.
		

\providecommand{\noopsort}[1]{}\providecommand{\singleletter}[1]{#1}%

\end{document}